\documentclass[acmsmall]{acmart}

\usepackage{todonotes}
\usepackage{soul}
\usepackage{amsmath}

\graphicspath{{figures/}}

\begin{document}

\title{Theory and Practice of Algorithm Engineering}

\author{Jan Mendling}
\authornote{Corresponding Author}
\email{jan.mendling@wu.ac.at}
\orcid{0000-0002-7260-524X}
\affiliation{%
  \institution{Humboldt-Universit{\"a}t zu Berlin}
  \streetaddress{Unter den Linden 6}
  \postcode{10099}
  \city{Berlin}
  \country{Germany}
}

\author{Benoit Depaire}
\email{benoit.depaire@uhasselt.be}
\affiliation{%
  \institution{Hasselt University}
  \streetaddress{Martelarenlaan 42}
  \postcode{3500}
  \city{Hasselt}
  \country{Belgium}}

\author{Henrik Leopold}
\email{Henrik.Leopold@the-klu.org}
\affiliation{%
  \institution{K{\"u}hne Logistics University}
  \streetaddress{Großer Grasbrook 17}
  \postcode{20457}
  \city{Hamburg}
  \country{Germany}
}

\renewcommand{\shortauthors}{Mendling et al.}
\newcommand{\suggest}[1]{\sethlcolor{green} \hl{#1}}
\newcommand{\del}[1]{\st{#1}}
\begin{abstract}
\textbf{Abstract:} There is an ongoing debate in computer science how algorithms should best be studied. Some scholars have argued that experimental evaluations should be conducted, others emphasize the benefits of formal analysis. We believe that this debate less of a question of either-or, because both views can be integrated into an overarching framework. 
It is the ambition of this paper to develop such a framework of algorithm engineering with a theoretical foundation in the philosophy of science. 
We take the empirical nature of algorithm engineering as a starting point. Our theoretical framework builds on three areas discussed in the philosophy of science: ontology, epistemology and methodology. In essence, \emph{ontology} 
describes algorithm engineering as being concerned with algorithmic problems, algorithmic tasks, algorithm designs and algorithm implementations. \emph{Epistemology} describes the body of knowledge of algorithm engineering as a collection of prescriptive and descriptive knowledge, residing in World 3 of Popper's Three Worlds model. \emph{Methodology} refers to the steps how we can systematically enhance our knowledge of specific algorithms. In this context, we identified seven validity concerns and discuss how researchers can respond to falsification.
Our framework has important implications for researching algorithms in various areas of computer science. 
\end{abstract}


\maketitle
\citestyle{acmauthoryear}

\section{Introduction}
\label{sec:introduction}


The design and evaluation of algorithms have been a major concern of computer science since its founding days and is still a matter of discussion, as the recent debate for more~\cite{DBLP:journals/cacm/Mitzenmacher15} and less~\cite{DBLP:journals/cacm/Ullman15} experimental evaluations of algorithms shows. It was the ambition of \citet{dijkstra1968constructive} to make computing a field of mathematical inquiry that could help to prove correctness of algorithms. \citet{knuth1974computer} responded that the practical act of creating an algorithm is equally important. Later, \citet{sanders2009algorithm} added that the practical application aspect of algorithms should also be considered. On purpose, \citet{sanders2009algorithm} uses the term algorithm engineering, and not algorithm theory, to emphasize that algorithms have to empirically demonstrate their usefulness in real-world settings. In essence, the statements of these scholars reflect the tension between different views of how much research on algorithms is a matter of formal science or empirical science.

In this paper, we follow up on this debate. It is not our aim to engage in a normative discussion to which extent algorithms should be evaluated theoretically or empirically. We believe both positions have their merit and research in both areas provides valuable insights. Instead, we develop a conceptual argument that both views fit into a single ontological, epistemological and methodological framework describing algorithm engineering. Our development of such a framework is important for research on algorithms and connected with the following challenges. 

As a first challenge, we observe that a review of the practice of algorithm engineering is missing. Much of methodological considerations seem to be implicitly clear within the domains of major conferences such as the Very Large Data Bases (VLDB) or the Visualization (VIS) conferences. On the one hand, it might be argued that such implicit standards reflect a consensus in the field of computer science, which can be interpreted as a sign of maturity of the field. \citet{kuhn2012structure} calls this \emph{normal science} to signify the incremental work in an established, consensual paradigm. On the other hand, the requirements for an evaluation in conferences like the mentioned VLDB and VIS are quite different with VIS putting a strong emphasis on user studies and VLDB on performance evaluation and formal correctness. There are good reasons for these differences, but is there a common scientific ground? Our framework provides an answer.

As a second challenge, we observe that many areas of algorithm engineering have turned to competitive testing with the help of benchmark data sets with established quality metrics to tease out the last epsilon of improvement. As already \citet{hooker1995testing} pointed out, such competitive testing provides knowledge about which algorithm is better with respect to specific accuracy metrics, but limited understanding of why. Still, benchmarking is an established approach in areas such as image processing~\cite{liang2015encoding} or computational linguistics~\cite{kwiatkowski2019natural}. The extensive iterative use of the same benchmark data has been criticized for overfitting the data, but it is hard to consider alternative avenues if authors do not have the meta arguments at hand to explicate that their deviation from common practice provides sound insights. Our framework is meant to help authors tailoring their research designs to such different needs.  

We approach the identified challenges by developing a theoretical framework that integrates different perspectives on algorithms and their evaluation from computer science, operations research, design science, and the philosophy of engineering. This framework describes an \emph{ontological} perspective of what algorithm engineering is concerned with, an \emph{epistemological} perspective of what knowledge algorithm engineering aims to formulate, and a \emph{methodological} perspective of how algorithm engineering advances our understanding of algorithms. 
In this way, we clarify that different performance propositions are connected with different types of evaluation. 

This article is structured as follows. Section~\ref{sec:background} discusses the theoretical background of our framework for algorithm engineering. Section~\ref{sec:ontology} focuses on the ontological perspective of our framework. Section~\ref{sec:epistemology} describes its epistemological perspective. Section~\ref{sec:methodology} discusses the methodological perspective. 
Finally, Section~\ref{sec:conclusion} concludes our paper.


\section{A Theoretical Framework for Algorithm Engineering}
\label{sec:background}
 \emph{Algorithm Engineering} is concerned with algorithms and engineering.
In this context, an \emph{algorithm} is understood as a well-defined sequence of computational steps that transforms some input into some output~\cite{cormen2009introduction}. Algorithms can be understood as a specific class of technological rules~\cite{bunge1967search}. A key property of algorithms is that the number of steps, the effort they require and the time they take has to be finite \cite{aho1974design}.

\citet{staples2014-ontology}, with reference to \citet{vincenti1990engineers} and \citet{rogers1983nature}, defines \emph{engineering} as the practice of organizing the design, construction and operation of any artifact that transforms the world around us for meeting a recognized need. \citet{staples2014-ontology} emphasizes three aspects of this definition. First, engineering deals with artifacts. In our context, we focus on algorithms as specific artifacts. Second, these artifacts are meant to meet specific requirements. Third, engineering builds on theories and propositions that organize knowledge on why artifacts meet requirements. In this way, they help to explain and predict behavior or, more specifically, the performance of an algorithm in a particular setting.

Upon this basis, we can define \emph{algorithm engineering}. \citet{sanders2009algorithm} summarizes the historic evolution of the term from its first mentioning in 1980s and initiatives in the 1990s on developing libraries of efficient algorithms for research and reuse. With reference to \citet{demetrescu2004algorithm}, he defines algorithm engineering as the \emph{discipline that focuses on the design, analysis, implementation, tuning, debugging, and experimental evaluation of algorithms. }
The salient feature of this definition is the term \emph{discipline}. This term indicates that the design of algorithms is not the end of algorithm engineering, but the phenomenon that is studied. This means that its end 
is the generation of scientific knowledge about algorithms and their design. In this regard, algorithm engineering and algorithm theory are actually alike. They both use scientific methods to produce scientific knowledge with respect to algorithms. They do, however, differ in their method and the nature of the knowledge they uncover. 
While algorithm theory largely builds on the formal sciences, algorithm engineering puts more emphasis on empirical evaluation of algorithms using experiments.
Both \citet{staples2014-ontology} and \citet{sanders2009algorithm} agree that engineering and algorithm engineering, respectively, are empirical disciplines to which principles of falsifiability defined by \citet{popper1959logic} apply. This means that theories of algorithm engineering build on \emph{empirical theories}, which in contrast to mathematical theories make falsifiable propositions about real-world phenomena related to algorithms and their performances. 

To understand the importance of algorithm engineering, it is useful to reflect on the limitations of algorithm theory. From the perspective of algorithm theory, implementing algorithmic ideas is considered part of application development \cite{sanders2009algorithm}. However, since the early 1990s it became apparent that there are many reasons why this stance leads to a growing gap between theory and practice. Most notably, modern hardware with parallelism, memory hierarchies, etc. significantly differs from traditional machine models. Hence, theoretical performance considerations are not always practically meaningful \cite{sanders2009algorithm}. The fact that two algorithms differ with respect to their theoretical complexity class does not necessarily mean that their performance also differs on two real-world data sets. \citet{kriegel2017black} also demonstrated the impact of implementations in this context. They showed that many algorithmic improvements are actually achievable with optimized implementations and do not require true algorithmic improvements. Another important limitation of algorithm theory is its focus on efficiency \cite{sanders2009algorithm}. In practice, an NP-complete algorithm might still be very useful because it can effectively support a human in performing a complex task. However, obtaining insights into these aspects requires the use of empirical methods from algorithm engineering.  

Against this background, we take the empirical nature of algorithm engineering as a starting point. Our goal is to develop a theoretical framework along three dimensions, each related to a sub-discipline of the philosophy of science: ontology, epistemology and methodology. In essence, \emph{ontology} is concerned with ``what is''. For algorithm engineering, this means we need to clarify what are the phenomena that we consider. \emph{Epistemology} deals with the nature of knowledge about a specific phenomenon - i.e. ``what can we know about algorithms''. \emph{Methodology} refers to the study of method. It is concerned with the question ``how can we systematically enhance our knowledge'' of specific algorithms. 

In the following Sections~\ref{sec:ontology}--\ref{sec:methodology}, we discuss each of these three sub-disciplines for algorithm engineering. 
We illustrate our argument with reference to different examples of algorithms from sorting, shortest-path identification, process model discovery, and word-sense disambiguation~\cite{navigli2009word}. In short, sorting algorithms efficiently transform a list into an ordered list~\cite{aho1974design}. Shortest-path algorithms take as input a list of places that have to be visited and provide as output an optimal shortest path between them~\cite{deo1984shortest}. 
Process model discovery algorithms take event sequence data as input and generate a (typically simplified) visual representation using a process model~\cite{DBLP:books/sp/Aalst16}. Word-sense disambiguation algorithms map a word to its most likely sense in the context of a sentence or text~\cite{navigli2009word}.


\section{Ontology of Algorithm Engineering}
\label{sec:ontology}
Ontology is the study that is concerned with the structure of the real world~\cite{wand1990ontological}. An ontological theory of engineering therefore clarifies the structure of real-world phenomena that relate to engineering.
We build upon the ontological model of \citet{staples2014-ontology} and refine it for algorithm engineering. This model identifies aspects in a precise and explicit way according to principles generally accepted in computer science \cite{aho1974design} and operations research \cite{landry1983model}.
In essence, our ontological model approaches algorithm engineering as a process. The activities of this process are concerned with transitioning from one aspect to the next:
\begin{enumerate}
    \item The starting point is an \emph{algorithmic problem}, which resides in a specific usage situation and specifies the need to move from an as-is state towards a desired to-be state. 
    \item This algorithmic problem is analyzed and translated into an \emph{algorithmic task}. This task conceptualizes the algorithmic problem and captures the essential assumptions and requirements.
    \item Next, the analyst constructs an \emph{algorithm design}. This design addresses the algorithmic task and incorporates design principles and design decisions.
    \item Finally, the implementation transforms the algorithm design into an \emph{algorithm implementation}. This implementation reflects implementation decisions and materializes the design. The resulting program can be executed on data to address the original algorithmic problem.  
\end{enumerate}

 \begin{figure}
 	\centering\includegraphics[width=0.35\linewidth]{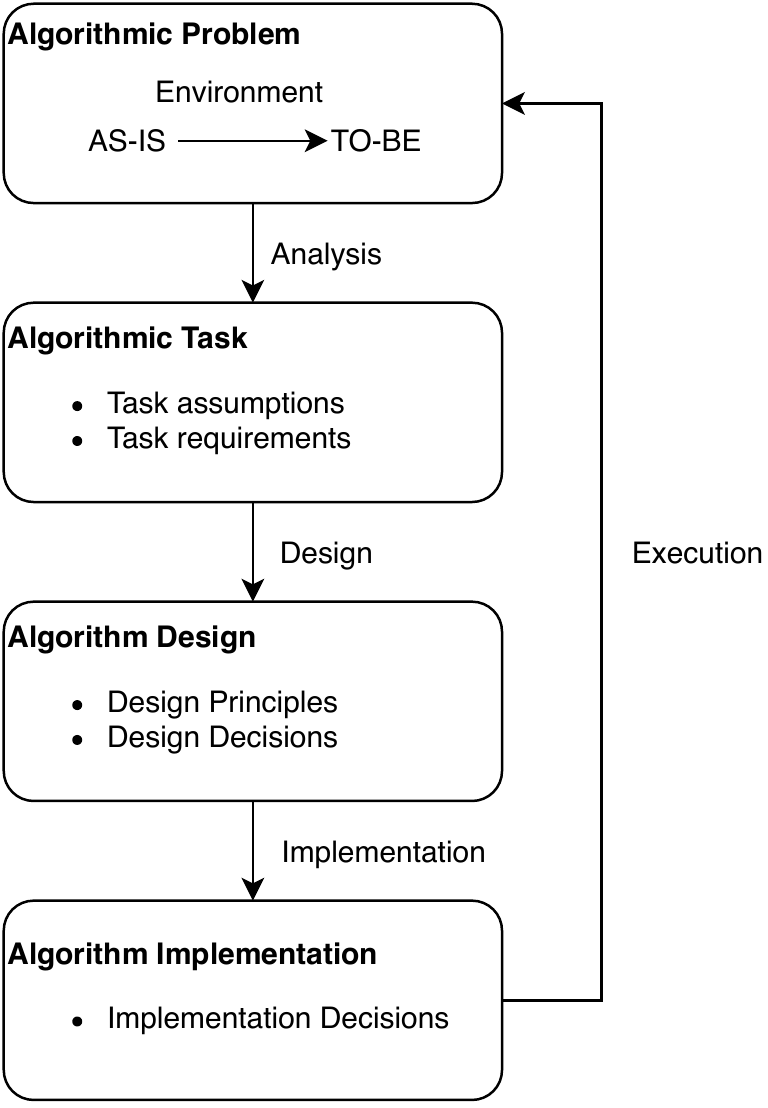}
     \caption{A framework for algorithm engineering: the ontological perspective}
 	\label{fig:ontology}
 \end{figure}
Figure~\ref{fig:ontology} provides an overview of our ontological framework. We discuss its four aspects in turn.

\subsection{Algorithmic Problem}

An algorithmic problem is embedded into a particular usage situation. This problem is the motivation for conducting algorithm engineering research. Different terms have been used for referring to the usage situation. \citet{meyer2015nested} speak of a problem domain. The usage situation in engineering is characterized by a difference between an \emph{observed state} of the world and a \emph{desired state} of the world. \citet{hall2017design} refer to this difference as a need, while~\citet{DBLP:books/sp/Wieringa14} speaks of goals and desires. More broadly, we refer to the difference between an observed state and a desired state as an \emph{algorithmic problem} if an algorithm can help to transition from the observed towards the desired state. 

\subsection{Algorithmic Task}

Many algorithmic problems expose themselves as deeply intertwined with the specifics of a particular usage situation. These specifics establish themselves as singular problems. The abstraction of the problem helps to translate the algorithmic problem into an algorithmic task. This implies that an algorithmic task corresponds to a class of algorithmic problems. The more abstract an algorithmic task is formulated, the broader its extension of real-world problems it subsumes. It is through this very act of problem abstraction that the ambition of science to formulate knowledge about \emph{classes of problems}~\cite{aken2004management} is supported. Over time, this has resulted in a variety of well-known and newly-established problem classes, such as the sorting problem, automated process model discovery~\cite{DBLP:books/sp/Aalst16}, or the word-sense disambiguation problem~\cite{navigli2009word}. Extensive taxonomies of problem classes exist, such as for shortest-path algorithms~\cite{deo1984shortest}.

Problem abstraction is the act by which the analyst translates the algorithmic problem into assumptions and requirements. Assumptions are associated with the as-is state and the environment of the algorithmic problem. They 
are outside the control of the algorithm. Requirements on the other hand, are associated with the to-be state and set out the goals for the algorithm. With respect to a shortest-path problem, the algorithmic task can be defined with the assumption that the input is a weighted graph without negative edges together with two requirements: find the shortest path in the graph between a set of given vertices and do so in a minimum amount of time. Note that a whole class of specific algorithmic problems are covered by this algorithmic task. 

In general, tasks vary in terms of complexity and clarity. If tasks are crisp and all required information is locally available, then these tasks can be fully automated by the help of algorithms \citep{sedlmair2012design}. Sorting belongs to the category of these tasks. Tasks can also be fuzzy where information partially resides in the head of the analyst. Then, interactive techniques can be designed that integrate several algorithmic components \citep{meyer2015nested}. This is often the case for tasks in computer visualization research. Also process mining tools integrate process model discovery algorithms with interactive support for abstraction and filtering. In this way, the user can, for instance, decide to exclude rare events or event sequences from the resulting process model. Word-sense disambiguation is a fuzzy task since the semantics of natural language can hardly be fully specified. Such fuzzy tasks 
are often characterized by uncertainty, conflicting options, and incomplete information \cite{campbell1988task}. Ultimately, a key challenge for the analyst is to specify the algorithmic task in such a way that an algorithmic design is feasible.

\subsection{Algorithm Design}
The act of designing an algorithm refers to a specific algorithmic task. Whereas the algorithmic task essentially describes \emph{what} is to be done and in which context, the algorithm design specifies \emph{how} it is to be done. The act of design takes the assumptions and requirements described by the algorithm task and translates them into abstract data structures and a design specification that meets the desired quality properties~\cite{DBLP:books/sp/Wieringa14}. 
%
Design is guided by knowledge~\citep{staples2014-ontology} on why algorithm designs meet specific requirements. Designers of algorithms make use of solution strategies for devising a novel algorithm. Several generic strategies have been identified. For instance, \citet{aho1974design} describe individual algorithms as specific instantiations of generic algorithmic strategies like divide-and-conquer, dynamic programming, greedy algorithms, backtracking, or local search. \citet{blum2003metaheuristics} discuss so-called meta-heuristics, such as tabu search or ant colony optimization, as strategies to define heuristic methods that can be applied to a wide range of different problems. Such generic solution strategies are often more generally applicable than for just one algorithmic problem. They are referred to as \emph{design principles}~\cite{gregor2013positioning}.

An algorithmic design also incorporates \emph{design decisions}, either explicitly or implicitly. These are specific for the task at hand. The actual integration of a design principle into the algorithm design is an example of such an explicit decision. Also the parameterization of certain aspects of design are explicit design decisions; they are a means to postpone the decision. Some decisions are more implicit and have subtle implications. For instance, when the algorithm is specified using Pascal-like pseudo code, this implies the assumption that it will be implemented using a procedural programming language. This entails also assumptions about the machine that is going to execute the programming code~\cite{aggarwal1988input,sanders2009algorithm}. 

The design, its design principles, and the corresponding design decisions are made explicit in a design specification, which abstracts from individual programming languages. There are various formats for explicating an algorithmic design, ranging from a declarative specification using mathematical formulae, a procedural specification using Pascal-like pseudo code, towards an informal description using structured text with nested enumerated lists of operations. The specification in whatever format defines the sequence of operations that constructs the output from the given input data.

\subsection{Algorithm Implementation}

Through the act of implementation, a developer produces an algorithm implementation. This implementation is meant to instantiate the algorithm design. The implementation closes the circle: it is the concrete solution to the original algorithmic problem. It can be executed on a computer to address the specifics of a given usage situation.

In practice, the implementation is actually hardly ever a simple instantiation~\cite{lukyanenko2020design}. Depending on the implementational choice of a programming language, it might be the case that operations of the algorithmic design cannot be directly translated to constructs of the target programming language. Furthermore, the runtime execution of the programming code inherits characteristics of the respective execution environment including hardware and system software characteristics. These depend upon memory hierarchy~\cite{DBLP:journals/eatcs/Sanders04}, caching strategies~\cite{karedla1994caching}, or how the compiler makes use of parallel processing opportunities provided by the processor~\cite{wolfe1996high}. 

\section{Epistemology of Algorithm Engineering}
\label{sec:epistemology}
The previous elaborations emphasize that the practice of designing algorithms 
is concerned with finding an algorithm implementation for a given algorithmic problem. The science of engineering takes a more abstract view. It is not about solving individual problems, but about ``general knowledge, linking an intervention or artefact with a desired outcome or performance
in a certain field of application'' \cite{aken2004management}. 
Therefore, the goal of algorithm engineering 
is to extend the body of knowledge of algorithm designs, algorithmic tasks, and corresponding propositions.

The \emph{epistemological} perspective of our framework addresses questions of what we can know about an algorithm and when we can consider this to be valid knowledge. This is best made visible by describing how knowledge relates to design.

\begin{figure}
 \centering\includegraphics[width=0.8\linewidth]{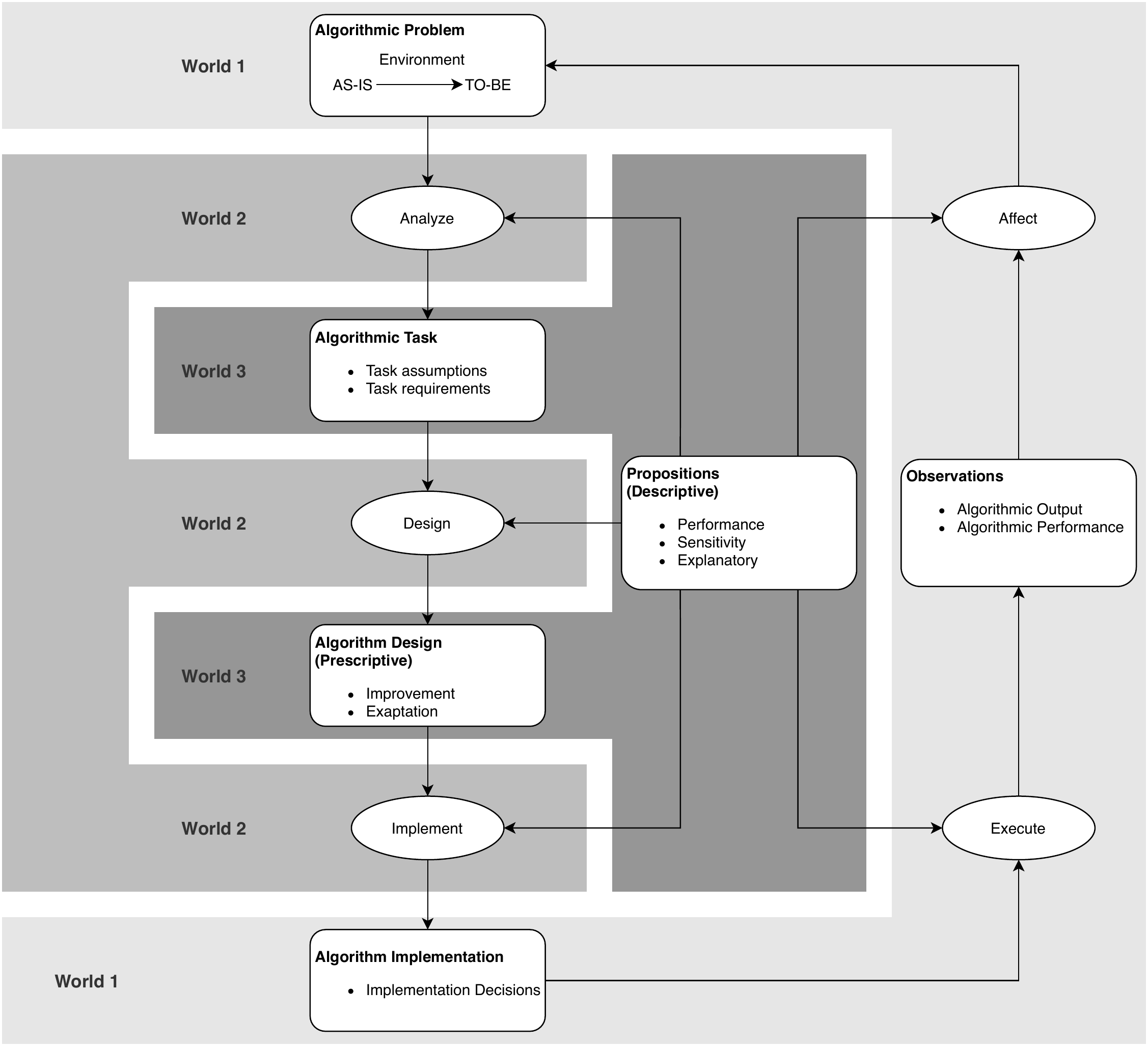}
 \caption{A framework for algorithm engineering: the epistemological perspective}
 \label{fig:epistemology}
\end{figure}

\subsection{Popper's Three Worlds}

The body of knowledge of algorithm engineering refers to different spheres that \citet{popper1979three} describes in his epistemological model of three worlds. In essence, he distinguishes World 1 of physical entities, World 2 of subjective mental states, and World 3 of objective knowledge. World 1, often referred to as the real world, is assumed to exist independent from an observer and to behave in a regular way. The goal of science is to formulate objective empirical knowledge in World 3 about World 1. Such scientific knowledge is formulated by humans and each of their individual mental states defines World 2 that mediates between the other two worlds.

Figure~\ref{fig:epistemology} shows our ontological perspective on algorithm engineering and how it relates to Popper's three worlds. First, both algorithmic problems and algorithm implementations are entities that exist in the physical world independently from an observer; they belong to World 1. 
Second, the three actions analyze, design, and implement build on human cognition and mental models. For this reason, they reside in World 2. 
Third, the algorithmic tasks, algorithm design and propositions with respect to both are part of World 3 as these are meant to be objectively described. 
Typical for these World 3 artefacts is that they abstract from their World 1 counterparts.

Popper's three world view unveils an important implication for algorithm engineering that is easily overlooked. From an epistemological perspective, knowledge exists in World 3 and thus algorithm engineering refers to the body of knowledge about the algorithm task, the algorithm design, and corresponding propositions. Mind that our understanding of the specific algorithmic problem or the specific algorithm implementation are ontologically different as they reside in World 1. World 1 matters from a practical point of view, but World 3 is what we are interested in from a scientific point of view. What we call the \emph{body of knowledge} resides in World 3.

To illustrate this point, consider process model discovery. Assume the algorithmic problem in World 1 is the need to understand the behavior of a customer complaint handling process at a  company, given a data set describing activities executed in the process. Analysts will then use their mental models (World 2) to make abstractions, eventually defining this problem as an algorithmic task. This task could be to automatically discover an understandable process model representing the behavior of the complaint handling process from an event log, where each event in the event log corresponds to the completion of a specific activity of a specific process instance. This algorithmic task specification abstracts from the real world and resides in World 3. Scientific knowledge on this task such as described by~\citet{augusto2018automated} helps us to understand which algorithms are most suitable for the data at hand.
Then, an engineer can construct mental models (World 2) to address this algorithmic task, using design principles and other available algorithm knowledge to specify an algorithmic design, which could be based on, for example, the design of the alpha algorithm~\cite{DBLP:journals/tkde/AalstWM04}, an existing algorithm for process discovery. As the design specification abstracts from implementation details, it is a World 3 artefact. 
Based on the design, a developer can produce an algorithm implementation (World 1), e.g. partially reusing libraries of existing open-source frameworks for process model discovery, such as ProM~\cite{DBLP:books/sp/Aalst16} or bupaR~\cite{janssenswillen2019bupar}, which both implement the same design, but make different implementation decisions. This actual algorithm implementation can be used to tackle algorithmic problem. 

\subsection{Algorithms and Popper's World 3}
The body of knowledge of algorithm engineering includes both knowledge \emph{of} algorithm designs as such and knowledge \emph{about} algorithm designs and their properties in relation to algorithmic tasks. This body of knowledge can be understood as a collection of \textit{algorithm designs}, \textit{algorithmic tasks}, and corresponding \textit{propositions}. 
\citet{staples2014-ontology} with reference to~\citet{simon1969sciences} defines the structure of propositions in engineering along three terms: the purpose or goal of an artefact, its character, and the environment in which it operates. Similarly, \citet{DBLP:books/sp/Wieringa14} defines propositions in terms of \emph{assumptions about context}, \emph{specification of the artefact} and the \emph{effect of the artefact}. \citet{DBLP:journals/infsof/HallR17} define a software problem in relation to an environment and a specific need of a problem owner. While terminology differs, there is consensus that propositions are constructed around three major concepts: the goal of the algorithm, the algorithm design and the algorithm's environment. These, in turn, correspond to the task requirements, the design decisions and the task assumptions in our ontological framework in Figure~\ref{fig:ontology}. 

Research aims for an extension of the body of knowledge. If such an extension is supported by a valid justification, we call it a contribution. 
The spectrum of contributions ranges, a.o., from \emph{descriptive} types such as analysis, explanation, and prediction to \emph{prescriptive} types~\cite{gregor2006nature,johannesson2014introduction}. 
In relation to algorithms, two kinds of corresponding contributions can be made: design contributions, which are prescriptive, and knowledge contributions, which are descriptive.

\subsection{Design Contributions}
First, \emph{design contributions} present new combinations of algorithm tasks and designs. Design contributions are \emph{prescriptive}. \citet{gregor2013positioning} refer to them as $\Lambda$ knowledge, representing design theories that are emerging. For instance, design principles, strategies, technological rules, methods, or heuristics belong to this category. Software design patterns~\cite{gamma1995design} and metaheuristics~\cite{blum2003metaheuristics} are prominent examples. \citet{gregor2013positioning} distinguish in this context:
\begin{enumerate}
    \item Design Improvement: a better performing algorithm design is presented for an established task, and
    \item Design Exaptation: an established algorithm design is adapted for a new type of task. 
\end{enumerate}
On the one hand, established tasks like sorting are subject to continuous efforts of improvement, such as e.g.~\cite{abdel2017efficient}. Here, the problem space is fixed and new algorithms expand the solution space. 
On the other hand, generic algorithm designs such as deep learning~\cite{lecun2015deep} are continuously adapted to new tasks. Here, the algorithmic solutions are transferred to new kinds of problems in an act of exaptation. 
Both are combined in what \citet{gregor2013positioning} specifically call invention.

\subsection{Knowledge Contributions}
Second, \emph{knowledge contributions} present new propositions and corresponding support about algorithm designs and their properties. This knowledge is descriptive.
\citet{gregor2013positioning} speak of $\Omega$ knowledge that refers to well-developed design theories. 
Insights into algorithms are neither arbitrary nor self-evident. They require a foundation in either novel or reference theories. \citet{DBLP:books/sp/Wieringa14} speaks of knowledge context. The benefits of theory development have been emphasized in various empirical branches of computer science such as software engineering~\cite{johnson2012s,wohlin2015general,DBLP:journals/tse/Ralph19} and computer visualization~\cite{sedlmair2012design}.
In this context, various kinds of knowledge contributions have been discussed~\citep{staples2015-methodology,karl1963conjectures}.

Following work by~\citet{santner2003design} on the design and analysis of computer experiments, we distinguish three types of propositions that are common in the field of algorithm engineering. These cover the non-prescriptive spectrum of contributions~\cite{gregor2006nature,johannesson2014introduction}:
\begin{enumerate}
    \item Performance Propositions: These propositions focus on the algorithm's performance in terms of the extent that the algorithm \emph{meets its task requirements}. Performance propositions range from determining the expected performance of an algorithm, whether the algorithm meets a specific threshold or that an algorithm outperforms state-of-the-art algorithms. Research focused on performance propositions often relies on a competition-based evaluation or formal proofs. 
    \item Sensitivity Propositions: These propositions relate to robustness and uncertainty of the performance of an algorithm. Robustness propositions make statements \emph{how robust} the performance of an algorithm is in face of changes of \emph{internal} design decisions. These design decisions are often parameterized and the goal is to evaluate the robustness of the algorithm's performance against sub-optimal parameter settings. 
    Uncertainty propositions consider the performance of an algorithm as a function of the task assumptions, which describe the relevant aspects of the \emph{external} environment of an algorithm. The goal of uncertainty propositions is to evaluate the uncertainty of the expected algorithm performance as the algorithm environment varies across problem instances. Alternatively, the goal could also be to investigate the precision of propositions about the expected performance.
    \item Explanatory Propositions: Explanatory propositions provide insights into the \emph{mechanisms} of how task assumptions and design decisions interact and influence the algorithm's performance. Two types of explanatory propositions can be distinguished. Weak explanatory propositions are typically derived from observational data and focus on associative patterns between task assumptions, design decisions and the algorithm's performance. Strong explanatory propositions describe causal factors and typically require an experimental setting or formal proofs. They identify the mechanisms how design decisions influence algorithm performance given specific task assumptions.
\end{enumerate}

In this way, knowledge contributions are structurally similar to design contributions. They either extend the coverage of the phenomenon space or the explanation space, while design contributions either enhance the problem space or solution space. 
Research papers differ in their emphasis of design and knowledge contributions. There are papers providing a design contribution together with a knowledge contribution. A proposal of a new algorithm is a typical example of this kind, where properties of the algorithm are analyzed to justify its underlying propositions. There are also papers providing mainly a knowledge contribution. Survey papers or benchmarking studies belong to this category.

\subsection{Contributions Exemplified}
For illustration purposes, consider the vehicle routing problem class with time windows (VRPTW). A VRPTW defines a routing problem of a vehicle fleet with known capacity that serves a set of clients with known demands from a central depot and where each client needs to be served within a specific time window. The goal is to minimize the total cost or distance traveled.

In~\citet{pisinger2007general}, an adaptive large neighborhood search (ALNS) framework is presented to tackle the VRPTW and the authors provide performance and uncertainty propositions with respect to their algorithm. As for the \emph{performance propositions}, the authors initially provide some graphs that show the performance of the algorithm on a specific problem instance and how the solution cost drops over time as the algorithm runs. Based on these results, the authors claim that the algorithm behaves very typically for a simulated annealing metaheuristic. This is a reference to \emph{prescriptive knowledge}. Next, the authors evaluate the performance of their algorithm against a set of state-of-the-art algorithms across a well-known benchmark of VRPTW problem instances, leading to the proposition that the ALNS framework is able to compete with the best heuristics for the VRPTW. Ultimately, the authors evaluate \emph{uncertainty propositions} whether performance scales when the number of clients to serve (task assumption) increases. They proposition support that their approach is among the two heuristics that reach the best solution quality.

In~\citet{corstjens2019}, the authors consider a variant of the ALNS framework and create an experimental setup, which allows them to run a regression model of the performance on the design decisions. These design decisions are both design principles such as specific local search heuristics that have been implemented or not into the ALNS framework and parameters that guide the ALNS framework. Based on the results of the regression analysis, the authors can formulate both \emph{robustness propositions} as well as \emph{explanatory propositions}. They identify a set of design decisions that do not have a significant effect on the algorithm's performance, hence claiming robustness. Additionally, the authors study the interaction between the destroy heuristics and the repair heuristics employed in the ALNS framework and reveal patterns how these design decisions interact and correlate with the algorithm's performance. These weak explanatory propositions are further analyzed experimentally in a related study, which support strong explanatory propositions on the mechanisms behind these patterns~\cite{corstjens2018explaining}.

While the ontological framework describes the different concepts and the epistemological framework describes what we can know and upon which basis. What remains is the question how we establish new knowledge on algorithms. This is addressed by the third part of our framework, the methodology of algorithm engineering. 

\section{Methodology of Algorithm Engineering}
\label{sec:methodology}
\begin{figure}
 \centering\includegraphics[width=0.8\linewidth]{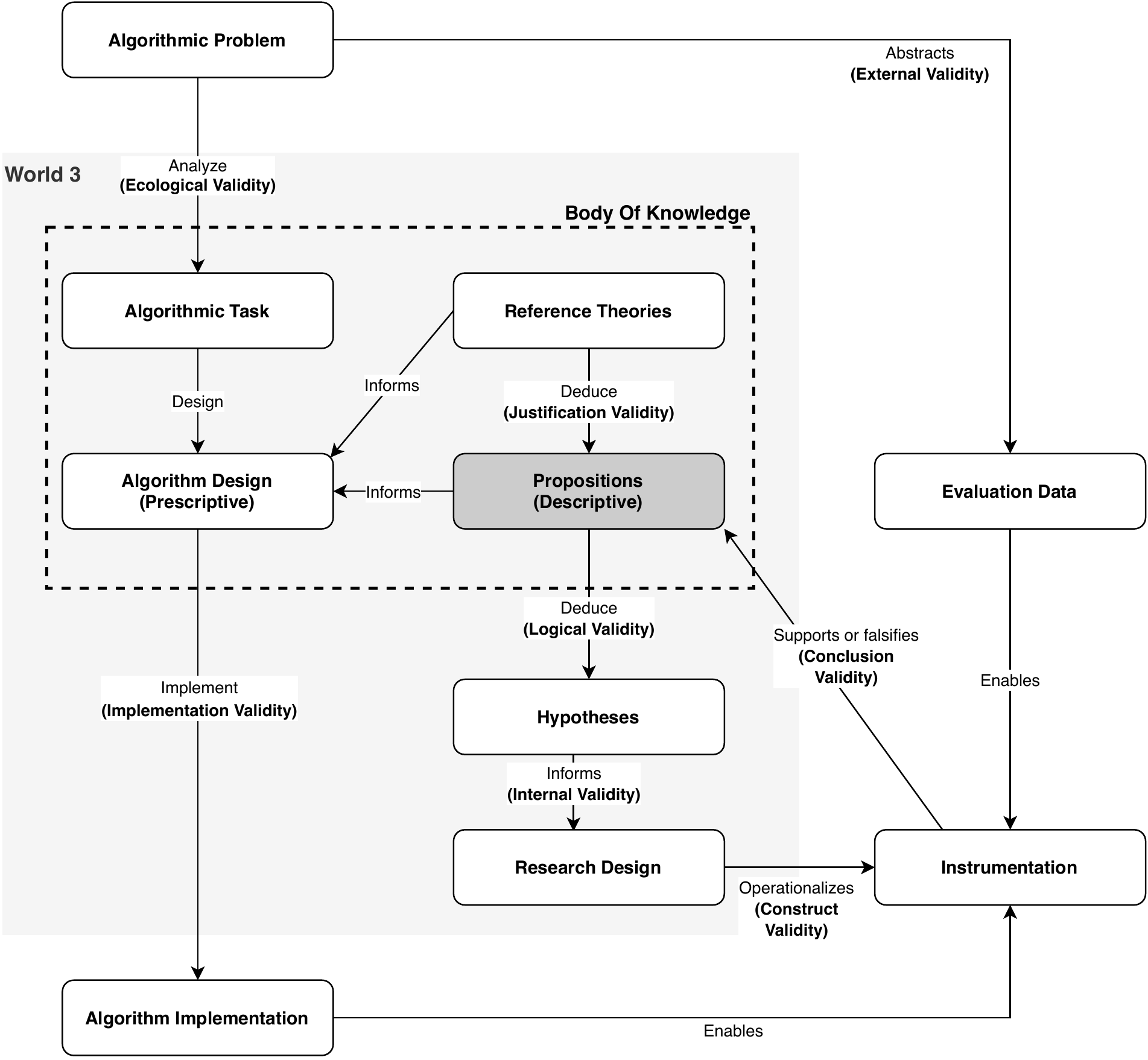}
 \caption{A framework for algorithm engineering: the methodological perspective}
 \label{fig:methodology}
\end{figure}

We emphasized above that the general goal of algorithm engineering is to extend the body of knowledge of algorithm designs, algorithmic tasks, and corresponding propositions. Up to this point, however, we have focused on \textit{what} we can know. In this section, we elaborate on the question of \textit{how} to systematically extend the body of knowledge. More specifically, we focus on how to systematically establish knowledge contributions, i.e, supported propositions with respect to the algorithm design.

Figure \ref{fig:methodology} summarizes the methodological perspective of our framework. 
This perspective focuses on the generation and validation of \emph{new} algorithm knowledge, which is at the center of an iterative process~\cite{sanders2009algorithm,DBLP:books/sp/Wieringa14}.
New propositions require a justification that is often constructed with the help of reference theories. Propositions are then the basis for devising a research design and its corresponding hypotheses. This research design is then operationalized into experimental instrumentation. These experiments build on the implementation of the algorithm design and evaluation data, and lead to a number of observations. These observations might either support the propositions or lead to a revision of these propositions or the algorithm design. In the former case, the propositions can be considered as new validated \textit{algorithm knowledge} that enhances the body of knowledge. 

In the following, we describe the details of the methodological perspective. Our specific focus is on validity threats. To this end, we distinguish six types of validity that relate to the justification of propositions, research design, operationalization, implementation, use of data and conclusions. 
Finally, we discuss responses to falsification.  

\subsection{Justification of Designs and Propositions}
Algorithm designs and corresponding propositions are neither arbitrary nor self-evident. They require a foundation in either prior or novel theory. \citet{DBLP:books/sp/Wieringa14} calls this knowledge context. 
The benefits of theory development have been emphasized in various empirical branches of computer science such as software engineering~\cite{johnson2012s,wohlin2015general,DBLP:journals/tse/Ralph19} and computer visualization~\cite{sedlmair2012design}.
Algorithm designs and propositions about them have been justified by theories drawn from different scientific disciplines. These categories are also referred to as \emph{reference theories}.

\begin{description}
    \item[Computer Science:] First, computer science, and in particular, algorithm theory provides insights into algorithms based on mathematics and logic. The theory of computational complexity~\cite{papadimitriou2003computational} provides a mathematical foundation for deriving runtime boundaries for algorithmic efficiency.
    Second, computer science preserves a collection of established algorithm designs for reuse and refinement. The books by~\citet{DBLP:books/aw/Knuth68,DBLP:books/aw/Knuth69,DBLP:books/aw/Knuth73} belong to the most prominent examples of this category.
    \item[Natural Sciences:] Optimization and research on metaheuristics partially builds on phenomena studied in the natural sciences, such as simulated annealing~\citep{van1987simulated}, particle swarm optimization~\citep{kennedy1995particle}, or backpropagation for artificial neural networks~\citep{werbos1994roots}.
    Cognitive theories of human computer interaction~\cite{preece2015interaction}, theories of engineering psychology and human performance~\cite{wickens2015engineering} help judging the usefulness of output generated by specific algorithms. 
    \item[Social Sciences:] Information systems research~\cite{gregor2013positioning}, web science~\cite{hendler2008web}, and theories on the organizational impact of information technology~\cite{markus1983power,orlikowski1992duality} describe how algorithms and information systems can have an impact on organizations.
\end{description}

A key concern for the algorithm design and corresponding propositions in this context is \emph{justification validity}. This validity concern is a matter of subsumption of the phenomena associated with the considered algorithm.
Subsumption essentially builds on a deductive scheme, but comes with specific challenges. \citet{alexy2003balancing} highlights that the potential applicability of alternative rules is always possible, such that balancing is required. In this case, alternative propositions might be possible that are supported by different theoretical rules. Identifying plausible alternative propositions~\citep{DBLP:journals/tse/Ralph19} is therefore that foundation for balancing and justifying that the developed proposition is the most plausible one.

\subsection{Hypothesis Formulation}
New propositions about algorithms can be derived inductively or deductively~\cite{recker2012scientific} and they generally build on foundations established by reference theories. 
In turn, hypotheses are deductively constructed from propositions. Hypothesis formulation according to principles of falsification require the statement of a null hypothesis and an alternative hypothesis~\cite{wohlin2012experimentation}. The null hypothesis states that empirical connections are random while the alternative hypothesis states the opposite. The objective is to reject the null hypothesis, which provides support for the alternative hypothesis (or only ``hypothesis'' for short).

As an example, consider a novel algorithm design for a sorting problem. Let us assume the hypothesis is that the novel design yields a better performance than the state-of-the-art design. This means, we would like to substantiate a proposition based on a performance analysis. 

The key concern for hypothesis formulation is \emph{logical validity}. This validity concern is a matter of deduction from the presented propositions and its underlying theoretical justification. It refers to \emph{why} a hypothesis is supposedly valid. The backing of a theoretical justification is essential for establishing an explanation for the expected effect as stated by the hypothesis.

\subsection{Research Design}
The term research design refers to the plan used to examine a research question of interest~\cite{marczyk2005essentials}.
A central concern of the research design is \emph{internal validity}. A research design is internally valid when its manipulation is causally responsible for an observed effect. To this end, experimental designs build on randomization and blocking to eliminate the effect of potentially confounding factors~\cite{wohlin2012experimentation}. Randomization is a means to statistically eliminate confounding factors by randomly assigning units to treatment groups. Blocking eliminates confounding factors by keeping them at a constant level. For instance, consider a new algorithm is claimed to perform better than a recent alternative. Then, both can be run repeatedly on the same machine to avoid confounding effects of memory usage by randomization. Furthermore, both can be implemented in the same programming language by the same developer to block these factors~\cite{kriegel2017black}.

The most simple experimental design is a one-factor design with two or more levels. Often, the algorithm design is considered as this factor with several alternative algorithms representing the different levels. The effects of the factor can be evaluated using statistical tests of mean comparison such as ANOVA or non-parametric alternatives~\cite{wohlin2012experimentation}.
Correlational designs use data on various phenomenon-related properties without explicit control. Connections between the different data characteristics and performance can be assessed using correlation and regression.
Correlational studies have been more often used in empirical software engineering in the early 1980s, e.g.~\cite{basili1983metric}, but later largely replaced by experiments for better control of confounding factors and, accordingly, better internal validity.

\subsection{Instrumentation}
The instrumentation of the research design takes the implementation of the algorithm and evaluation data as input to produce performance measurements of the computational process and the generated output.

First, an \emph{implementation} of the algorithm design is used for the instrumentation. \citet{kriegel2017black} demonstrate that different implementations of the same algorithm design can vary in performance by orders of magnitude. A key concern here is~\emph{implementation validity} (or instantiation validity~\cite{lukyanenko2015guidelines,lukyanenko2020design}).
One threat for implementation validity is that the design generally leaves space for alternative implementation options. It is possible that implementation decisions turns out to be confounding factors~\cite{lukyanenko2015guidelines}. 
Another threat is a potentially unfaithful implementation. Testing cannot fully assure the conformance with the design specification. Optimization, debugging and publishing code helps to establish the desired confidence~\cite{kriegel2017black}. Also the formal analysis of the algorithm design in terms of correctness, completeness and termination supports implementation validity.

Second, \emph{evaluation data} is used as input for the instrumentation. A key concern of evaluation data is \emph{external validity}, which relates to the problem of variability in the real world~\cite{DBLP:journals/scp/WieringaD15}. As we cannot cover that variability completely, 
we have to establish that the input data at least covers the relevant and realistic range of to-be-expected input~\cite{sanders2009algorithm}. As neither complete nor random sampling of the whole potential range of input data is feasible, it is a good practice to vary features related to the size and complexity of the input data~\cite{kriegel2017black} or, more generally, study features of the phenomena represented by the data~\cite{lhermitte2011comparison}. \emph{Real-world data sets} (so-called organic data) are often assumed to have such external validity, but various validity threats arise if it is unclear whether their features ranges are realistic~\cite{xu2019validity}. The key challenge is, however, that data do not uniquely identify the mechanisms by which they were generated~\citep{pearl2019seven}. As a consequence, real-world data likely reflects various overlaid factors, making it difficult to untangle these factors when analyzing the performance of an algorithm. 
\emph{Artificial} (or computer-simulated) data sets \cite{santner2003design} might be the better choice once relevant input data features are understood. Simulation affords the systematic variation of typical input ranges, which is instrumental for generating knowledge into how input data features translate into algorithmic performance variations. Simulated data is particularly useful when empirical data i) does not exist, ii) is insufficient, or iii) imprecise~\cite{hofmann2013ontologies}.
As an example, consider a linguistic parsing algorithm that tries to detect the grammatical structure of a sentence. One question with respect to the performance of this algorithm could be how well it can deal with typographical errors in a text. In an artificial data set, both the number as well as the kind of typographical errors can be controlled and varied, best according to types and distributions observed in practice. In this way, it is possible to obtain specific insights into the performance of this algorithm design in presence of typographical errors. 

In some communities, so-called \textit{benchmark data sets} are publicly shared to encourage the performance evaluation of competing algorithms. They can stem from real-world domains or artificial simulation. Among others, such benchmark data sets are available for part-of-speech tagging \cite{paul1992design}, image recognition \cite{fei2006one}, ontology matching \cite{algergawy2019results}, process model matching \cite{cayoglu2013report}, and vehicle routing problems \cite{DEFRYN2016400}. Benchmark data sets not only facilitate comparative evaluation. They also play an important role for exploring how specific features of realistic input data can be exploited. \citet{sanders2009algorithm} points to cases where benchmarking data inspired drastic algorithmic improvements. 
It is important to note that the use of benchmark data sets does not necessarily mitigate threats to external validity. Also a benchmark data set might have a limited scope or lead to overfitting over time \cite{sim2003using,tichy1998should}, which essentially means that not algorithms, but research teams building algorithms are compared~\cite{hernandez2017evaluation}.

Third, many propositions around algorithms refer to \emph{measures of performance}, either of the computational process or the quality of the generated output. The measures used in the instrumentation have to be valid and reliable operationalizations of the propositions~\cite{DBLP:books/daglib/0084392}. This is in essence a question of \emph{construct validity}, i.e.\ whether a measure of a construct sufficiently measures the intended property \cite{o1998empirical}. 
This challenge is equally relevant for user assessment of output quality as for execution time measures, which seem objective at a first glance. \citet{kriegel2017black} call for caution with wall-clock time. Measures also have to be relevant. For the linguistic parsing algorithm, the notion of performance is typically operationalized using the number of true positive, false positive, true negative, and false negative classifications \cite{hossin2015review}. The most commonly employed metrics in such a classification setting are precision, recall, and F-measure (the harmonic mean between precision and recall). The relative importance of these measures depends on the specific algorithmic problem. For a typical web search algorithm, we would be mainly interested in recall and less in precision. 
For a full text search algorithm, we might be mainly interested in the execution time. For an algorithm performing automated text summarization, we may require actual human feedback (e.g. assessing the usefulness of the output on a likert scale). These examples illustrate that, to a large extent, the type of algorithmic problem and the specific task requirements determines what measure is suitable for the experiments.

Finally, the results obtained via the measurement can either support or falsify the propositions. The key concern here is \emph{conclusion validity}, i.e.\ to which degree the conclusions we draw from the analysis of the results are reasonable for the propositions~\cite{cook1979design}. While the term conclusion validity strongly emphasizes statistical analysis, it also covers qualitative considerations \cite{cozby2007methods}. As an example, consider an experimental research design that yields as a result that the average runtime of a new algorithm design is faster than the state-of-the-art. Concluding that the new algorithm design is generally faster might, however, still be a unreasonable conclusion. This is, for instance, the case if the standard deviation is very high and, accordingly, the mean differences are not significant. In such a setting it might still be reasonable to conclude that the algorithm design is faster under certain conditions. Additional analysis of data points that exhibit anomalies or outliers can help to support such conclusions~\cite{hernandez2017evaluation} or to revise design or propositions.

\subsection{Responding to Falsification}
The \emph{comparison} of the expected with the observed results offers either support or a failure to reject the null hypothesis. In the latter case, it is challenging to identify the cause of the deviation. \citet{staples2015-methodology} describes potential reasons for failures. First, failures can stem from a mismatch between algorithm problem and implementation in World 1. A wrong implementation of the design is in this category. Second, failures of understanding can be associated with World 2 and any mental activity of interpretation and analysis. To address this point, \citet{kriegel2017black} recommends a fair comparison of algorithms by having them implemented by the same developers or developers with comparable expertise.
Third, failures can go across Worlds: algorithm task versus problem, algorithm design versus task, or algorithm implementation versus design. Algorithm problems reside in World 1 and are often ``wicked problems''~\cite{rittel1973dilemmas}. Their wicked nature is the reason why algorithm tasks are not an isomorphic mapping of the problem; they have to be constructed. Therefore, the concern of \emph{ecological validity} is of particularly importance, i.e.\ to what extent the findings of a scientific study can be generalized to a real-life setting \cite{brunswik1943organismic}.

In essence, there are two responses to a failure to reject the null hypothesis~\cite{staples2015-methodology}. We can revise the algorithm task, design, or implementation (design revision) or we revise the propositions, hypotheses, research design, instrumentation, or the underlying theory (knowledge revision). In this way, scientific knowledge creation manifests itself as an iterative process that is directed towards aligning the algorithmic problem and solution spaces, and the corresponding knowledge claims.

\section{Conclusion}
\label{sec:conclusion}
In this paper, we presented a theoretical framework of algorithm engineering. 
To this end, we took the empirical nature of algorithm engineering as a starting point. Our theoretical framework builds on three areas discussed in the philosophy of science: ontology, epistemology and methodology. In essence, \emph{Ontology} 
describes algorithm engineering as being concerned with algorithmic problems, algorithmic tasks, algorithm designs and algorithm implementations. \emph{Epistemology} describes the body of knowledge of algorithm engineering as a collection of prescriptive and descriptive knowledge, residing in World 3 of Popper's Three Worlds model. \emph{Methodology} refers to the steps how can we systematically enhance our knowledge of specific algorithms. In this context, we identified seven validity concerns and discussed how researchers can respond to falsification.

Our framework has important implications for researching algorithms in various areas of computer science. First, it provides algorithm engineering a theoretical foundation in the philosophy of science. This foundation clarifies how algorithms can be studied empirically. Second, our framework provides guidance for the methodological evaluation of newly proposed algorithms. Authors can use considerations that are related to the different validity concerns as a starting point for designing their evaluations. Third, this methodological grounding offers a framework for conducting research on algorithms in various areas of computer science in a way that supports an incremental research path. Finally, we, the authors, would be glad to receive feedback on this paper.

\bibliographystyle{ACM-Reference-Format}
\bibliography{refs}

\end{document}